\documentclass{optica-article}
\journal{optica}
\articletype{Research Article}

\begin{document}

\title{Frequency stability of cryogenic silicon cavities with semiconductor crystalline coatings}

\author{
Dhruv Kedar,\authormark{1,*}
Jialiang Yu,\authormark{2}
Eric Oelker,\authormark{1,3}
Alexander Staron,\authormark{1}
William R. Milner,\authormark{1}
John M. Robinson,\authormark{1}
Thomas Legero,\authormark{2}
Fritz Riehle,\authormark{2}
Uwe Sterr,\authormark{2}
and Jun Ye\authormark{1}
}

\address{
\authormark{1}JILA, NIST and University of Colorado, 440 UCB, Boulder, Colorado 80309, USA\\
\authormark{2}Physikalisch-Technische Bundesanstalt, Bundesallee 100, 38116 Braunschweig, Germany \\
\authormark{3}University of Glasgow, Institute for Gravitational Research, School of Physics and Astronomy, Glasgow G12 8QQ, United Kingdom
}

\email{\authormark{*}dhruv.kedar@colorado.edu}

\begin{abstract}
State-of-the-art optical oscillators employing cryogenic reference cavities are limited in performance by the Brownian thermal noise associated with the mechanical dissipation of the mirror coatings. Recently, crystalline Al$_{1-x}$Ga$_{x}$As/GaAs coatings have emerged as a promising candidate for improved coating thermal noise. We present measurements of the frequency noise of two fully crystalline cryogenic reference cavities with Al$_{0.92}$Ga$_{0.08}$As/GaAs optical coatings. We report on previously unmeasured birefringent noise associated with anti-correlated frequency fluctuations between the polarization modes of the crystalline coatings, and identify variables that affect its magnitude. Comparing the birefringent noise between the two cryogenic reference cavities reveals a phenomenological set of scalings with intracavity power and mode area. We implement an interrogation scheme that cancels this noise by simultaneous probing of both polarization modes. The residual noise remaining after this cancellation is larger than both cavities thermal noise limits, but still lower than the instabilities previously measured on equivalent resonators with dielectric coatings. Though the source of these novel noise mechanisms is unclear, we demonstrate that crystalline coatings can provide stability and sensitivity competitive with resonators employing dielectric coatings.
\end{abstract}

\section{Introduction}

Ultra-stable lasers locked to cryogenic reference cavities represent an indispensable tool for rapid characterization of optical atomic clocks \cite{Oelker2019, SrI}, development of optical timescales \cite{Milner2019, Fortier2011}, and tabletop investigations of fundamental physics \cite{Kennedy2020, Mueller2003, Wiens2016}. The performance of state-of-the-art optical oscillators is fundamentally limited by length fluctuations associated with thermally driven mechanical dissipation within the mirror coatings \cite{Kessler2012, Numata2004, Hong2013}. Thus, low mechanical loss coatings with high reflectivity operating at low temperatures are desirable for improving the performance of these optical reference cavities \cite{Hagemann2014, Matei2017, Zhang2017, Robinson4K} and interferometric gravitational wave detectors \cite{aLIGO, ET, KAGRA2015, Voyager}.

\begin{figure}[h]
\centering
\includegraphics[width=8.2cm]{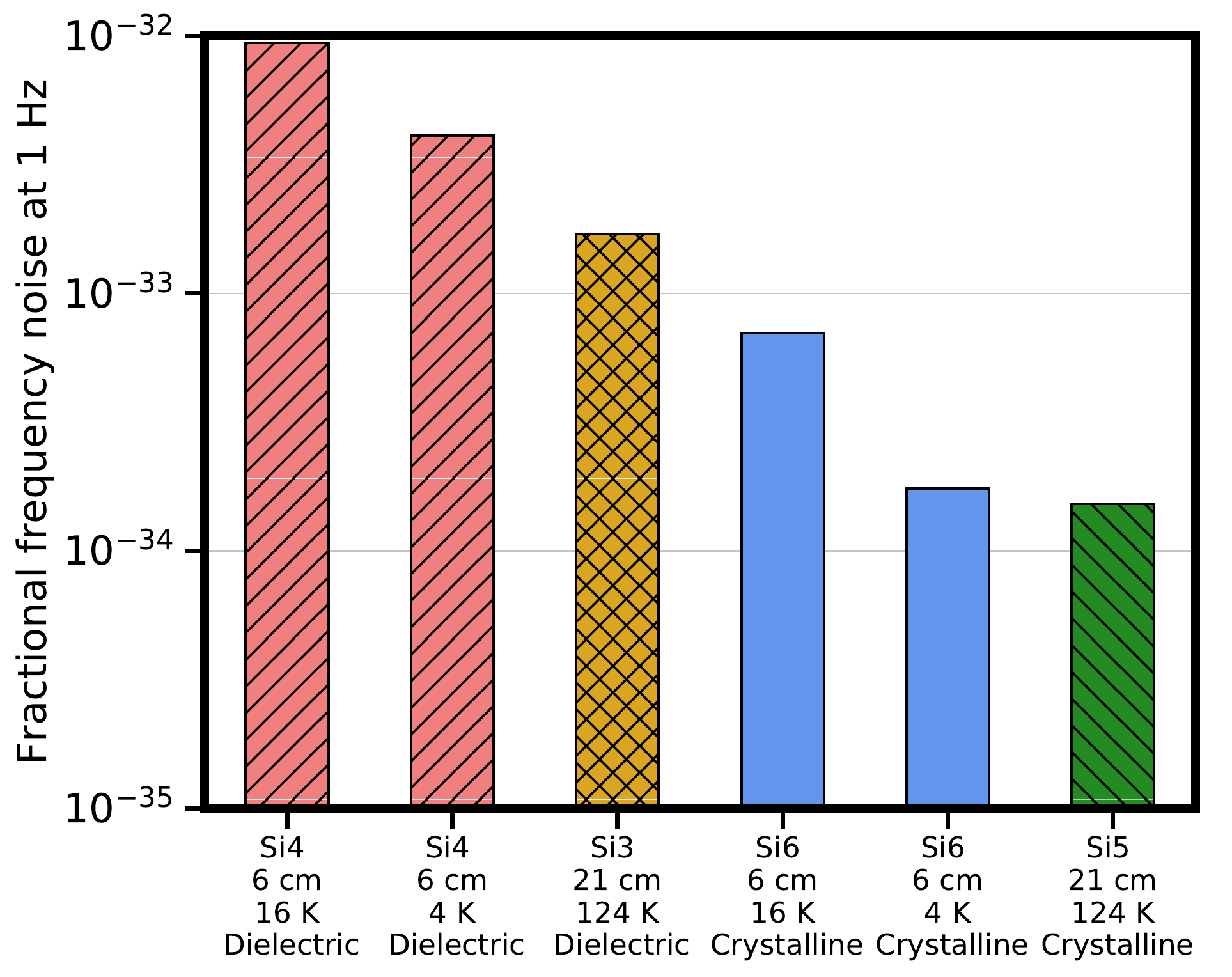}
\caption{Coating thermal noise contributions ($\approx S_{\textrm{Brownian}}$) for a variety of cryogenic silicon cavities referenced and compared in this work, with labels detailing spacer length, operating temperature, and optical coating material. The coating label "dielectric" specifically refers to SiO$_2$/Ta$_2$O$_5$, and "crystalline" refers to Al$_{0.92}$Ga$_{0.08}$As/GaAs. For thermal noise calculations of Si5 and Si6, we assume the room temperature value for the Al$_{0.92}$Ga$_{0.08}$As/GaAs loss angle \cite{Cole2013} is maintained at cryogenic temperatures. Conversely, Si3 and Si4 achieved thermal noise limited performance at these levels \cite{Oelker2019,Robinson4K,RobinsonCTN}.}
\label{fig:CTN}
\end{figure}

Studies of crystalline mirror coatings have revealed several materials that exhibit lower mechanical loss than their traditionally-used amorphous dielectric counterparts \cite{Cole2010, Cumming2015}. Room temperature loss angles of crystalline Al$_{0.92}$Ga$_{0.08}$As/GaAs ($\phi = 2.5\times 10^{-5}$ \cite{Penn2019}) are measured to be over ten times smaller than those of amorphous SiO$_2$/Ta$_2$O$_5$ layers (4 K loss angle of $5.6\times10^{-4}$ \cite{RobinsonCTN}). However, the majority of studies have extracted the coating loss angle via ringdown measurements of mechanical cantilevers \cite{Martin2010, Craig2019}, far from the frequency band of operation. Direct measurements of coating noise \cite{Cole2013} have been less frequently studied, but can provide broadband characterization of thermal noise and any other mechanisms that can generate sources of length noise. Interferometric measurements of coating noise are therefore imperative to the operation of gravitational wave detectors \cite{RobinsonCTN}. Furthermore, the associated length noise and loss angles have yet to be characterized in a cryogenic resonator system at the audio band and below. Cryogenic silicon cavities have been established as well-characterized and powerful platforms for studying the length noise associated with SiO$_2$/Ta$_2$O$_5$ dielectric optical coatings \cite{Kessler40mHz, Matei2017}. In this letter, we present the characterization of crystalline Al$_{0.92}$Ga$_{0.08}$As/GaAs coatings transferred onto silicon substrates and operated on a 6 cm monocrystalline silicon cavity spacer at temperatures of 4 K and 16 K, referred to as Si6. Using these coatings and assuming that the measured room temperature loss angle persists at our operating temperatures \cite{Cole2013}, we expect a fractional frequency instability of $1.3\times 10^{-17}$ at a cavity temperature of 4 K. A comparison is made to a previous system Si4 (Fig.\ref{fig:CTN}), an equivalent cavity with dielectric SiO$_2$/Ta$_2$O$_5$ coatings detailed in \cite{Zhang2017, Robinson4K}.

Investigations of the Si6 system reveal frequency noise associated with the large birefringent mode splitting of the crystalline mirror coatings. We identify an optical intensity scaling of this birefringent noise and compare it to measurements made on a similar 21 cm silicon resonator with crystalline coatings operated at 124 K (Si5). We suppress this noise by employing a novel simultaneous dual-tone probing of the two polarization eigenmodes of the cavity \cite{Hall2000}. The improved cavity stability on both systems is then limited by a residual noise of unknown origin, significantly above the expected thermal noise limit calculated using the room temperature loss angle of Al$_{0.92}$Ga$_{0.08}$As/GaAs.

\section{Methods}
Though the spacer and substrate parameters of Si4 and Si6 are identical, crystalline coatings introduce some key differences and complexities. The mirror coatings are composed of an 11.6 $\mu$m-thick multilayer stack of alternating high-index GaAs and low-index Al$_{0.92}$Ga$_{0.08}$As, and are oriented with parallel coating polarizations \cite{Cole2016}. The GaAs $\langle 100\rangle$ axis is normal to the coating, and parallel to the crystalline $\langle 111\rangle$ axis of the crystalline silicon. Two near-planar mirrors of 1 m radius of curvature realize cryogenic transmission ($T$) of 11.8 ppm and loss ($L$) of 10 ppm combined for both mirrors, supporting a finesse of 290,000. We measure a $770 \textrm{ kHz}$ birefringent splitting of the polarization modes, well resolved outside of the $8.6 \textrm{ kHz}$ cavity linewidth. The cavity is housed in a closed-cycle commercial cryostation that enables cooling the cavity down to a minimum temperature of 4.7 K.

\begin{figure}[t]
\centering
\includegraphics[width=8.2cm]{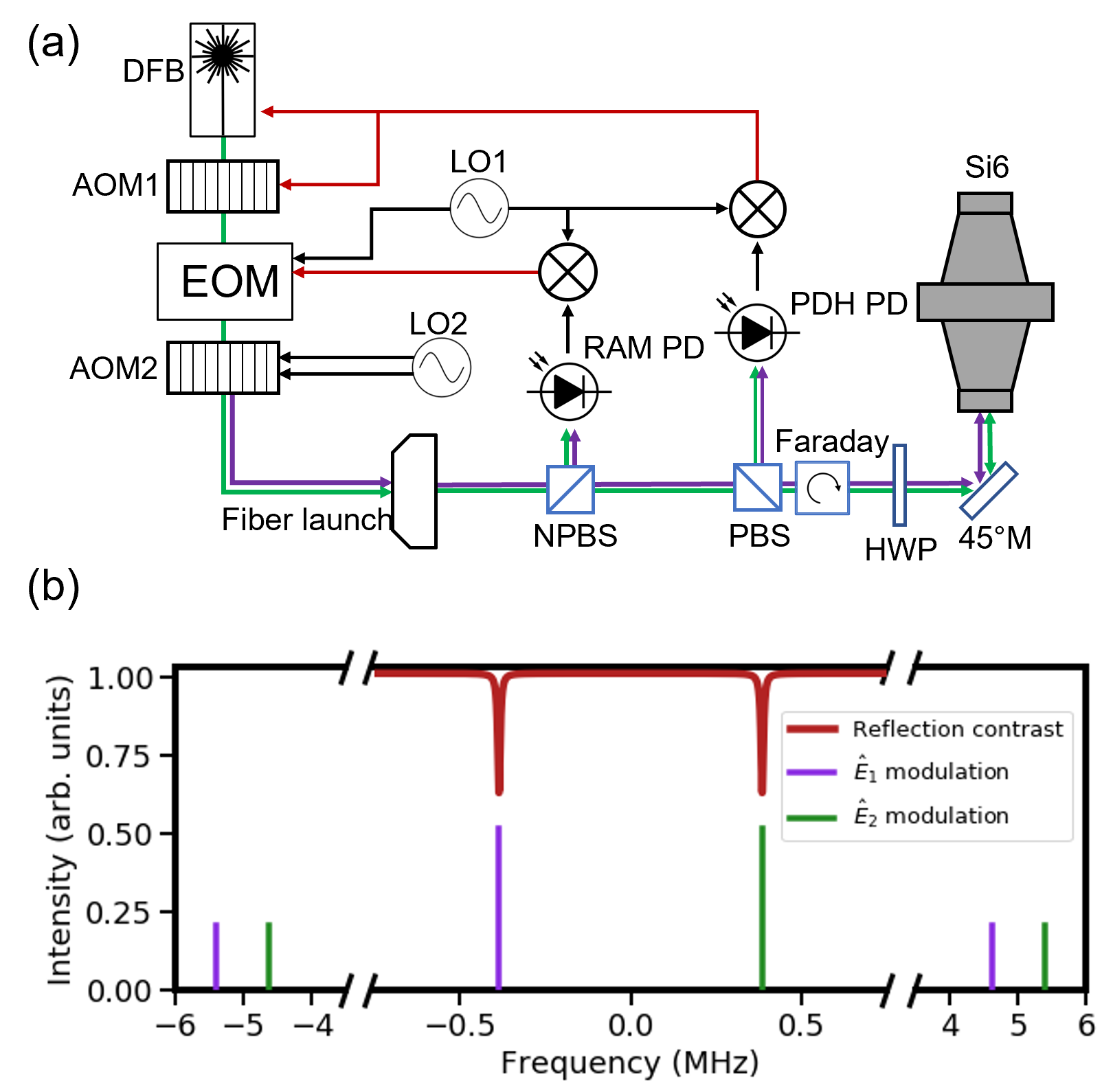}
\caption{(a) Optical layout for probing the Si6 system with active RAM suppression at a modulation frequency (5 MHz) where the cavity error signal is encoded in a traditional PDH locking setup. Dual-tone probing is achieved by driving AOM2 with two RF tones separated by the birefringent mode splitting of Si6 (green and purple sets of tones). A half waveplate before the cavity is aligned to couple an equal mixture of s- and p-polarized light to the two polarization modes of the cavity, labeled $\hat{E}_1$ and $\hat{E}_2$. The collinear beams are then launched upwards with a 45$^{\circ}$ mirror to couple into the cavity. The reflected cavity error signal has equal contributions from noise of the two polarization modes, and any anti-correlated noise is automatically rejected. (b) Frequency landscape of the tones applied to the Si6 cavity. The two tones applied to AOM2 separate the carrier and the two sidebands each into two tones separated by the birefringent mode splitting of the cavity. The purple tones shown address one polarization mode ($\hat{E}_1$) while the green tones simultaneously address the orthogonal mode ($\hat{E}_2$).}
\label{fig:cartoon}
\end{figure}

We interrogate the cavity with a unique measurement scheme that enables reduction of the frequency noise associated with the birefringent mode splitting. The optical layout is illustrated in Fig.\ref{fig:cartoon}(a). An error signal is derived from the cavity reflection by probing in the traditional Pound-Drever-Hall (PDH) technique, with modulation sidebands at 5 MHz generated via a fiber coupled electro-optic modulator (EOM). A portion of this signal is picked off before the cavity for active cancellation of residual amplitude modulation (RAM) with a control signal applied as a bias voltage to the EOM \cite{Zhang2014}. We stabilize the optical power in cavity transmission with feedback to an acousto-optic modulator (AOM1). Conventional PDH schemes address a single polarization mode of the resonator. This does not allow for any cancellation of birefringent noise, though it has not appeared to be a significant source of instability in cavities with dielectric coatings. Unique to this setup is the addition of AOM2 which synthesizes two frequency tones 770 kHz apart that can be used to address both polarization modes of the cavity. Both tones are imprinted with a corresponding set of phase modulation sidebands as illustrated in Fig.\ref{fig:cartoon}(b). A half waveplate before the cavity is set to couple an equal intensity of light into both polarization modes while maintaining equivalent discrimination slopes for their error signals. The detected signal on the PDH photodetector (PD) is the sum error signal of both polarization modes and any anti-correlated noise between the two is significantly suppressed. We routinely ensure that optical powers and discrimination slopes are matched to within $3\%$ between the two modes, and we project that this enables $> 20$ dB rejection of any anti-correlated signal. We note that RAM stabilization actuates on the sum error signals for both polarizations as well. Intensity stabilization suppresses any fluctuations in transmitted power for both optical modes. Si5 implements a different scheme for birefringent noise cancellation, detailed in \cite{Jialiang}.

\begin{figure}[t]
\centering
\includegraphics[width=8.2cm]{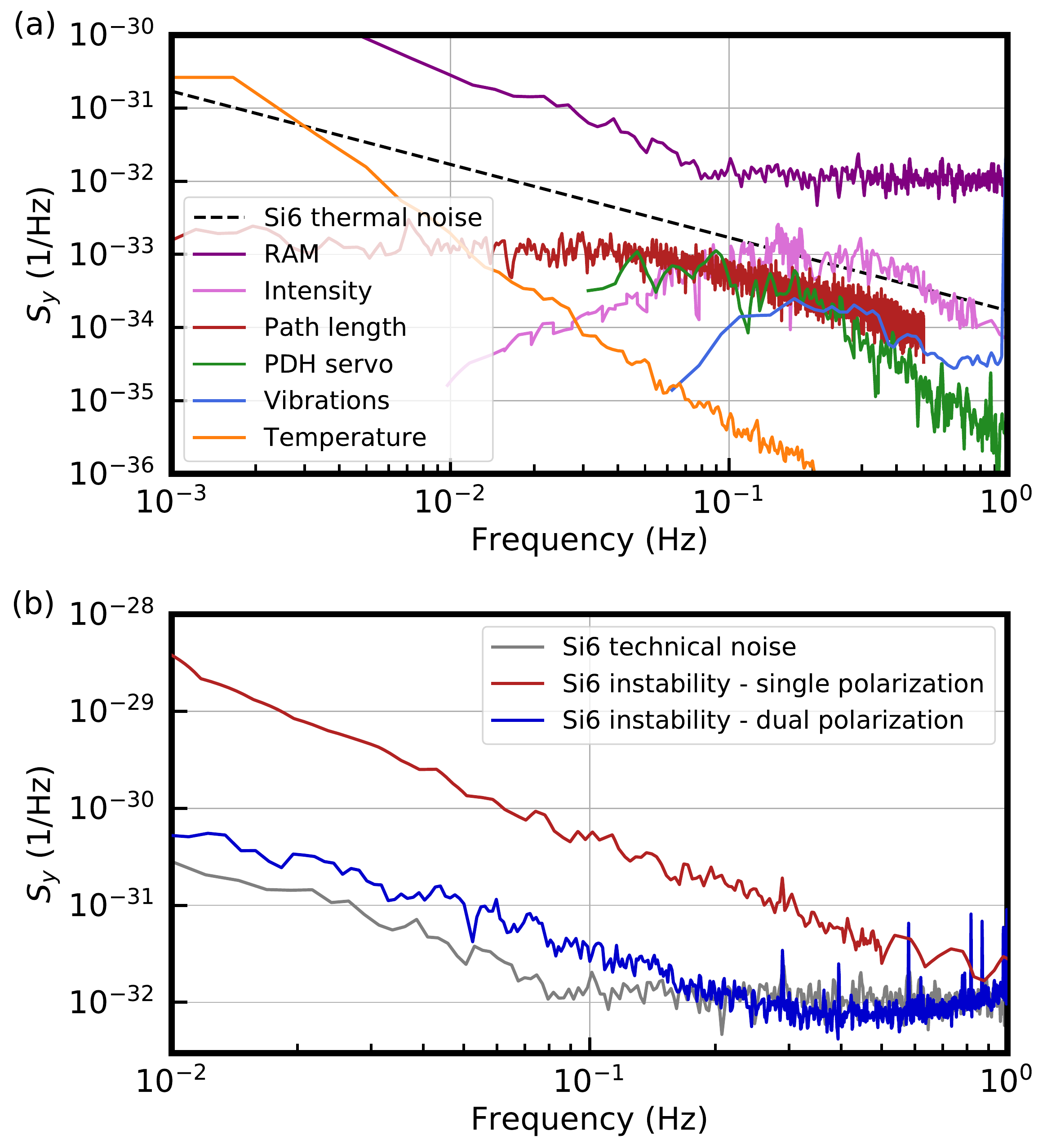}
\caption{(a) Technical noise contributions for Si6. The projected thermal noise is displayed (dashed black line). (b) Si6 frequency instability measured with single polarization probing (red trace, $P_c$ = 18 mW) and residual noise, with birefringent noise suppressed via dual polarization probing (blue trace). The red trace measures $(S_{\textrm{Brownian}} + S_{\textrm{biref}} + S_{\textrm{residual}})$, while the dual polarization lock only measures $(S_{\textrm{Brownian}} + S_{\textrm{residual}})$. A sum of technical noises is displayed as the grey trace.}
\label{fig:noisebudget}
\end{figure}

A compilation of characterized noise sources is detailed in Fig.\ref{fig:noisebudget}(a). Technical noise contributions are evaluated by propagating the ambient environmental noise through a measured transfer function. Significant upgrades were implemented to enable observation of the Si6 thermal noise. Noise contributions from electronic sources were reduced. The cavity mount was redesigned as a stainless steel tripod, and we identified a sinusoidal dependence of the horizontal acceleration sensitivities $\kappa_x, \kappa_y$ that oscillates with the azimuthal angle formed between the cavity $\langle 1 0 -1\rangle$ axis and the tripod support. Sinusoidal dependence of the horizontal acceleration sensitivity varies out of phase with the vertical sensitivity $\kappa_z$, and the operating angle is chosen to minimize the total vibration noise coupling to the cavity length. Ultimately we realize a 10-fold improvement over Si4 of horizontal acceleration sensitivity to values of $\kappa_x = 1\times 10^{-11} \textrm{g}^{-1}$ and $\kappa_y = 2\times 10^{-11} \textrm{g}^{-1}$, with vertical sensitivity $\kappa_z = 3\times 10^{-11} \textrm{g}^{-1}$ near DC.

The thermal environment of the cryostation, while conceptually similar to \cite{Zhang2017}, is also overhauled to support long term stability of projected Si6 thermal noise. All thermal shields are reconstructed out of high purity oxygen-free copper and gold-plated to reduce radiative coupling. The thermal isolation stages are engineered as a series of cascaded low pass filters. Adding holmium copper after the stage 2 coldfinger and integrating thermally resistive carbon fiber reinforced plastics into the support structure have increased the thermal isolation between cryogenic stages. The thermal time constant between active and passive 4 K thermal shields is increased nine-fold to 12000 s. The redesigned cavity support structure results in a reduced passive shield-cavity time constant of 600 s, down from 1200 s, despite superior vibration performance.

From Fig.\ref{fig:noisebudget}(a), it is clear that RAM dominates all other measured noise sources. We employ active RAM cancellation but are limited by out of loop parasitic effects occurring after the stabilization point (RAM PD in Fig.\ref{fig:cartoon}(a)). Aligning to the $\textrm{HG}_{00}$ cavity mode introduced an etalon with the input vacuum optic, and the etalon effect was strongly reduced by altering the angle of incidence to couple the $\textrm{HG}_{01}$ transverse mode. The reduced mode coupling efficiency increases the shot noise contribution, evident in the RAM signal.

Instabilities of Si6 are measured via direct beat over a noise-cancelled optical fiber \cite{Ma} with Si3, a 21 cm long silicon cavity with dielectric coatings operated at 124 K \cite{Oelker2019,Matei2017}. Knowledge of the reference laser's noise spectrum is crucial and the instability of the latter system is verified to be thermal noise limited with fractional frequency noise power spectral density $S_y = 1.7 \times 10^{-33}/f$ and fractional instability $\textrm{mod }\sigma_y = 3.5\times 10^{-17}$ from 1-1000 seconds \cite{Oelker2019}. The frequency noise of Si6 is calculated by measuring the combined fractional frequency instability of a Si3-Si6 optical heterodyne beatnote on a zero dead-time lambda-type counter. The Si3 instability contribution is then subtracted off in quadrature.

The other cryogenic silicon cavity employing Al$_{0.92}$Ga$_{0.08}$As/GaAs coatings (Si5) is 21 cm long and operated at 124 K \cite{YuEFTF, Jialiang}. Insight into the principle quantities driving the birefringent noise can be gained by comparing Si6 measurements to Si5. Coatings on this cavity share the same physical properties (layer stack, thickness) as those on Si6, though the optical qualities are slightly superior. Si5 realizes a finesse of over 350,000  corresponding to optical transmission and loss of $T = 11$ ppm, $L = 6.4$ ppm between both mirrors. The measured birefringent mode splitting of 205 kHz is roughly proportionally smaller due to the 3.5 times increased spacer length relative to Si6. Details of operation and technical noise characterization can be found in \cite{Jialiang}.

\section{Results}
We adopt the formalism of \cite{Jialiang} to describe the Brownian thermal noise, birefringent noise, and residual noise contributions to the measured instability of Si6:
\begin{equation}
    S_{\textrm{T}} = S_{\textrm{Brownian}} + S_{\textrm{biref}} + S_{\textrm{residual}}
\end{equation}

\begin{figure*}
\centering
\includegraphics[width=13.0cm]{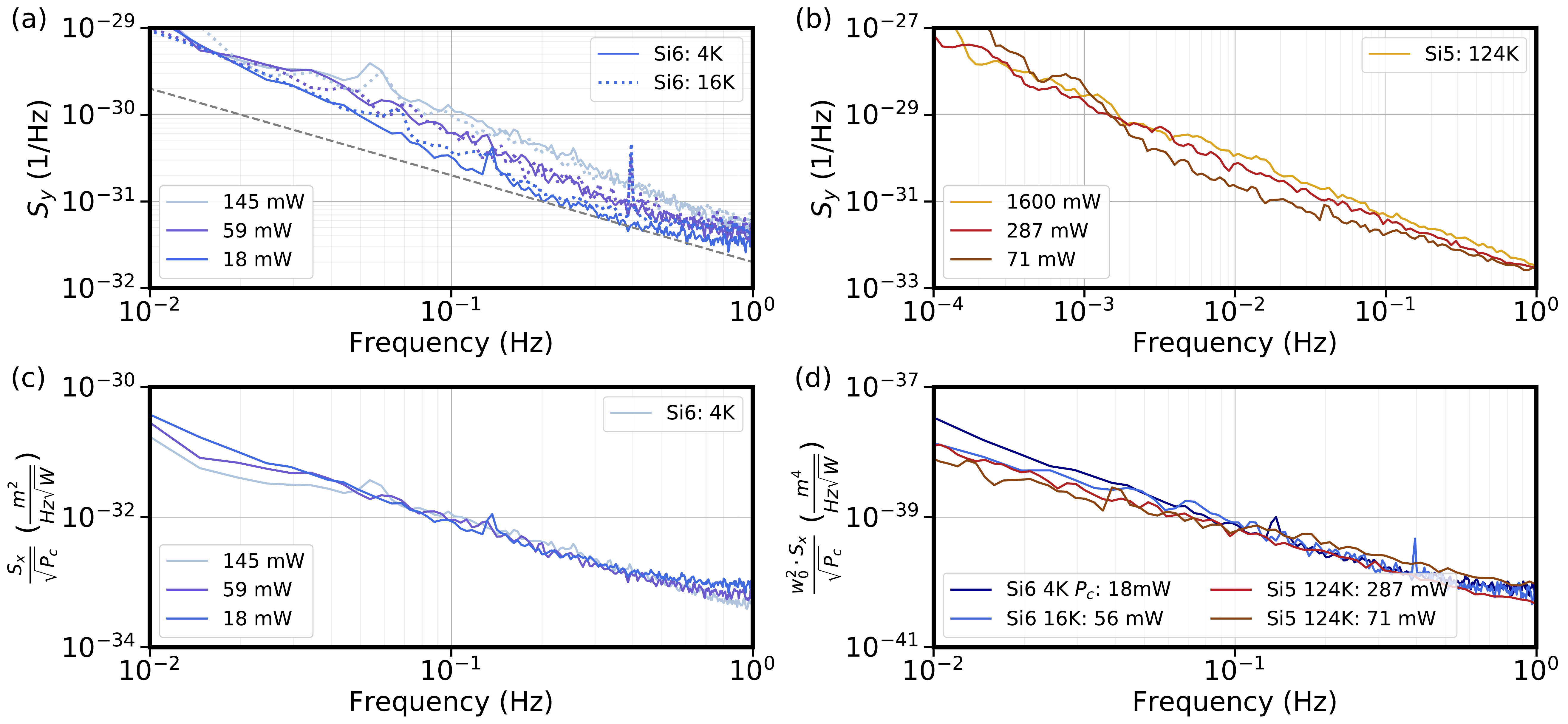}
\caption{(a) Birefringent noise measured with a single polarization lock exhibits a dependence on intracavity power $P_c$. Decreasing the intracavity power reduces frequency noise down to a limit of $P_c$ = 18 mW, where further reductions of power no longer result in an improvement of the cavity instability despite technical noise sources remaining well below the noise level. Although the character of the noise deviates strongly from $1/f$ in different frequency regimes, we include a bound of $2\times10^{-32}/f$ (dashed gray line) that identifies the lowest instability attained with the single polarization lock. (b) Power-dependent birefringent noise is also measured on Si5 at 124 K. (c) We observe an empirical scaling of this noise by $1/\sqrt{P_c}$ for both systems. Rescaling the length noise $S_x$ by this quantity yields a phenomenologically universal birefringent noise. As an example, we display Si6 data at 16 K with this rescaling implemented. (d) A comparison of the birefringent noise across the two systems, over a total of three different temperatures with four different intracavity powers. We rescale the length noise measurements by the observed dependencies on intracavity power and mode area and observe a universal behaviour of the birefringent noise.}
\label{fig:singledual}
\end{figure*}

On the Si6 system, the single polarization lock revealed that birefringent noise was significantly higher than all technical contributions (Fig.\ref{fig:noisebudget}(b)) as well as the projected coating thermal noise. Additionally, this birefringent frequency noise is observed to scale with the optical power coupled into the cavity, where a larger optical power corresponds to a larger cavity instability. The spectrum of this power dependent noise appears to exhibit a frequency-dependent slope which consistently exceeds a slope of $1/f$ in our measurement bandwidth. Below an intracavity power of 18 mW (100 nW measured in transmission), we do not measure any further reduction of birefringent frequency noise. We depict this with a dashed grey 1/f guide in Fig.\ref{fig:singledual}(a) corresponding to $S_y = 2\times10^{-32}/f$.

We increase the cavity temperature to 16.7 K, where silicon exhibits a zero crossing of the coefficient of thermal expansion \cite{Wiens14}. Comparison with the single polarization data taken at 4.7 K exhibits no difference, indicating that there is no strong thermal dependence of this phenomena between these two temperatures (Fig.\ref{fig:singledual}(a)).

\begin{figure*}[h!]
\centering
\includegraphics[width=8.2cm]{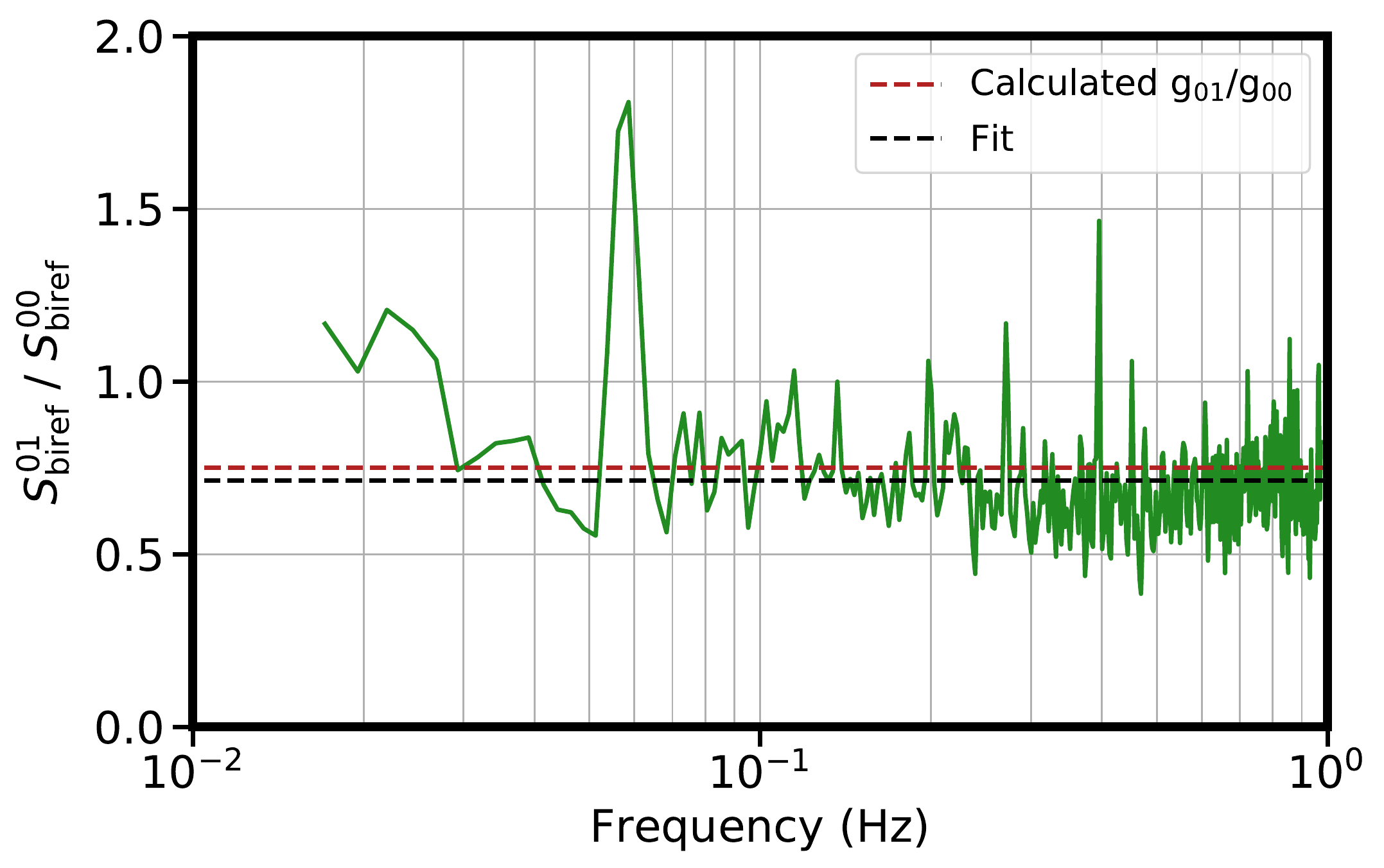}
\caption{Mode area effect on birefringent noise. We separately measure the birefringent noise of the $\textrm{HG}_{00}$ and $\textrm{HG}_{01}$ modes by addressing a single polarization of the cavity. The ratio of birefringent noise $S_{\textrm{biref}}^{01}$ and $S_{\textrm{biref}}^{00}$ for the two modes reveals the dependence on mode area. If the birefringent noise is independent of the mode area, we expect a ratio of 1. We find a ratio of 0.71, consistent with the expectation of 0.75 if the noise scales with mode area $g_{mn}w_0^2$. A large peak visible at 0.057 mHz is the result of a ground loop that appeared in the $S_{\textrm{biref}}^{01}$ dataset and was omitted in the fit.}
\label{fig:modearea}
\end{figure*}

We also find that the optical mode area on the mirror plays a significant role in the magnitude of the birefringent noise. Both cavities have interrogated the fundamental and higher Hermite-Gauss spatial modes $\textrm{HG}_{00}$ and $\textrm{HG}_{01}$ and find that $S_{\textrm{biref}}$ is reduced by the ratio of these two mode areas. We adopt the scaling factor $g_{mn}$ as defined in \cite{Vinet} for the reduction of coating thermal noise of higher order modes. When interrogating non-fundamental spatial modes, coating Brownian noise will be modified as $g_{mn}S_{\textrm{coating}}^{00}$ where $S_{\textrm{coating}}^{00}$ denotes the coating thermal noise of the $\textrm{HG}_{00}$ mode. For the higher $\textrm{HG}_{01}$ mode, the scaling factor is calculated as $g_{01} =0.75$. In Fig.\ref{fig:modearea} we measure the birefringent noise of the $\textrm{HG}_{00}$ and $\textrm{HG}_{01}$ modes at the same intracavity power and temperature by interrogating a single polarization in the Si6 system. Calculating the ratio $S_{\textrm{biref}}^{01}/S_{\textrm{biref}}^{00}$ will highlight any dependence on mode area. We find a ratio of 0.71$\pm 0.02$, similar to the expectation of $g_{01}/g_{00} = 0.75$ for birefringent noise dependent on mode area $w_0^2$.

Qualitatively similar noise characteristics for single polarization interrogation have been observed in Si5 \cite{Jialiang}. In contrast to measurements on Si6 where we infer the birefringent noise by observing the reduction of frequency with dual tone probing, birefringent noise in Si5 is measured directly by locking two independent lasers to the fast and slow polarization axes and measuring the frequency noise of their beatnote. Whereas $S_{\textrm{biref}} \approx S_{\textrm{residual}}$ in Si5, $S_{\textrm{biref}} \gg S_{\textrm{residual}}$  in Si6, so the inferred measurement of birefringent noise only contains a $25\%$ contribution of residual noise. Fig.\ref{fig:singledual}(a) and (b) display fractional frequency noise measurements made in both systems. Frequency noise $S_y$ is significantly lower on the 21 cm system for a given intracavity power, but we note that the length noise $S_x$ is the proper quantity to compare as this noise originates from the coatings. The longer spacer length of Si5 divides down the fractional length noise of the identical coatings, suggesting a possible strategy for mitigating birefringent noise. Comparison of $S_x$ between the two cavities provides significantly better agreement, identifying the crystalline coatings as the source for this excess birefringent noise. On both cavities we identify an empirical scaling of the birefringent with intracavity power as $1/\sqrt{P_c}$. That is, less power coupled into the cavity results in reduced length noise. Normalizing $S_x$ by this quantity (Fig.\ref{fig:singledual}(d)) produces a family of curves that appear to share some universal noise characteristics. 

The slope of this noise increasingly deviates from $1/f$ at lower frequencies. We note that the behaviour of this noise exhibits a similarity to electro-optic effects where the refractive index of a medium varies with an incident electric field. For Si5, an equivalent optical power bound cannot be identified. Birefringent noise at circulating powers of 127 mW is at a level comparable to the residual noise, thus a significant reduction of intracavity power provides a marginal reduction in measured instability. At this circulating power of 127 mW, we identify a frequency noise bound of $S_y = 1.9\times10^{-33}/f$ or $S_x = 8.5\times10^{-35}/f$. We stress that $\sqrt{P_c}$ scalings are empirical. Optical power build-up differs slightly between the two cavities and it is unclear whether circulating power or absorbed optical power is the more important metric. Nevertheless, rescaling by this quantity does appear to approximately capture the measured power dependence.

\begin{figure*}[h!]
\centering
\includegraphics[width=13.0cm]{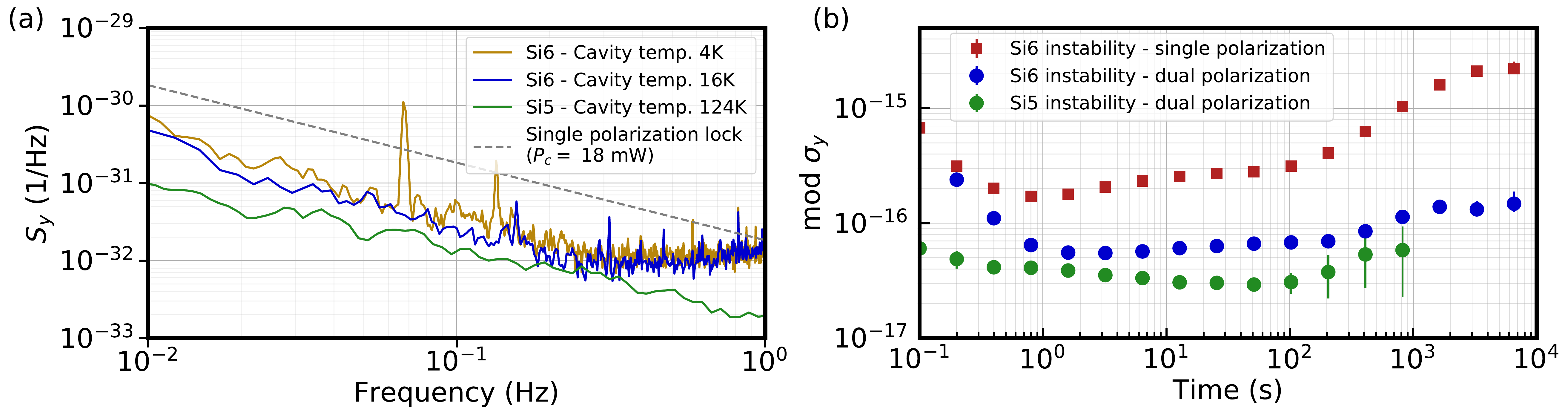}
\caption{(a) Residual noise $(S_{\textrm{residual}} + S_{\textrm{Brownian}})$ after cancellation of the birefringent noise at both 4 K and 16 K shows no temperature dependence. Residual noise of Si5 is at a similar level. (b) Long term instability of the residual noise of the two cavities, and birefringent noise of Si6. Both cavities exhibit significantly reduced long term instability when compared to their dielectric counterparts Si2 \cite{Matei2017} and Si4 \cite{Robinson4K}. Si6 realizes a fractional frequency instability of $5.5\times10^{-17}$, lower than the thermal noise limited instability of Si4. The $3.5\times10^{-17}$ fractional instability of Si5 matches the thermal limit of Si2.}
\label{fig:Si5Si6}
\end{figure*}

Unlike most fundamental sources of length noise, our measured birefringent noise does not appear to be a typical thermodynamically driven process. Investigating the noise of a single polarization mode at 4 K, 16 K, and 124 K, we do not identify a strong dependence on the temperature of the crystalline coating though data at all temperatures shares a similar scaling with optical power. Room temperature cavities employing crystalline coatings may therefore hide detrimental birefringent effects as the coating thermal noise can exceed birefringent noise for low intracavity power. 

A significant reduction in frequency noise is observed when we address both cavity polarization modes simultaneously by applying two RF tones to AOM2. We infer that the net improvement in cavity instability is due to first order cancellation of birefringent noise. The current level of instability is attributed to an unknown source of residual noise and no longer exhibits the intracavity power dependence observed with birefringence noise. This implies that the power dependence is a feature closely tied to the birefringent noise, and that the residual noise is not limited by imperfect cancellation of birefringence. Additional measurements where we introduce a slight imperfect cancellation of birefringence (i.e. improper matching of intracavity powers and discrimination slopes for the two modes) further confirms this, as the residual noise does not increase. 

Similarly, we find no dependence of the residual noise on which cavity spatial mode we interrogate. This contrasts to measurements of birefringent noise, where we observed that length fluctuations in the polarization modes splitting could be reduced by probing a larger area of the crystalline coating. Similar to studies in \cite{Jialiang}, we can establish that the residual noise is global in the sense that it is highly correlated over the length scale of the two spatial modes. It is possible that the current level of $S_{\textrm{residual}}$ is limited by an uncharacterized source of technical noise. In the frequency band of 0.01-0.3 Hz, we identify an instability consistent with $S_y = 3.37(9)\times 10^{-33}/f$ from a continuous 30,000 second measurement against Si3. This performance is slightly better than the thermal noise limited instability measured on Si4 of $S_y = 4.12(5)\times 10^{-33}/f$, and the long term instability for $\tau > 1000$ s significantly exceeds it. However, we note that the current instability of Si6 does not vary between cavity temperatures of 4.7 K and 16.7 K. This suggests that either coating thermal noise is not the primary source of instability, or that there is a strong temperature dependence of the AlGaAs/GaAs mechanical loss angle. 

Given that we find no reduction in instability when interrogating the larger HG$_{01}$ spatial mode, we infer that this residual noise is not the coating thermal noise limit. Using the formalism of \cite{RobinsonCTN} for calculating the coating thermal noise limit, we therefore place constraints on the maximum value of the coating loss angle $\phi$:

\begin{align}
    \phi_{4.7\textrm{ K}} < 4.1 \times 10^{-4} \\
    \phi_{16.7\textrm{ K}} < 1.2 \times 10^{-4}
\end{align}

These bounds already imply that these crystalline coatings can yield a sensitivity improvement over conventional dielectric coatings at 4.7 K, and a significant improvement at 16 K.

\section{Conclusion}
We have presented frequency noise measurements of crystalline Al$_{0.92}$Ga$_{0.08}$As/GaAs coatings measured on a 6 cm and 21 cm cryogenic silicon resonator. While the AlGaAs mechanical loss angle promises potential for improved coating thermal noise at 4 K, 16 K, and 124 K, we identify a novel source of instability associated with the mirror birefringence. We demonstrate that this birefringent noise is optical intensity dependent, does not depend strongly on temperature, and can be cancelled to reveal a temperature independent residual noise. Without cancellation, we note that the fractional frequency contribution of this birefringent noise can be reduced by employing longer resonators, but applications requiring large intracavity power may still be sensitive to these effects. The observed dependence on mode area suggests that increasing beam size can reduce the magnitude of the polarization fluctuations. Understanding the origins of this birefringent noise will be imperative for the development of low mechanical loss coatings as reduction of coating thermal noise is a significant obstacle for improving reference cavities and gravitational wave interferometers.
\\
\\
\noindent\textbf{Acknowledgments} The authors thank T. Brown and A. Ellzey for technical assistance, and G. Cole and J. Hall for insightful discussions.
\\ \\
\noindent\textbf{Funding} This work is supported by NIST, DARPA, AFRL, National Science Foundation Physics Frontier Center (NSF PHY-1734006), and Physikalisch-Technische Bundesanstalt. Ji.Y, T.L, F.R, and U.S. acknowledge support from Project 20FUN08 NEXTLASERS and from the Deutsche Forschungsge-
meinschaft under Strategy–EXC-2123 QuantumFrontiers, Project-ID 390837967; SFB 1227 DQmat, Project-ID 274200144.
\\ \\
\noindent\textbf{Disclosures} The authors declare no conflicts of interest.
\\ \\
\noindent\textbf{Data Availability} Data underlying the results presented in this paper are not publicly available at this time but may be obtained from the authors upon reasonable request.

\bibliography{refs}

\end{document}